\documentclass[11pt,a4paper]{article}
\usepackage[hyperref]{naaclhlt2018}
\usepackage{times}
\usepackage{latexsym}

\usepackage{multicol}
\usepackage{booktabs}
\usepackage{amsmath}
\usepackage{url}
\usepackage{bm}

\aclfinalcopy 

\def\R{\mathbf{R}}
\def\ReLU{\textnormal{ReLU}}

\providecommand{\tightlist}{%
  \setlength{\itemsep}{0pt}\setlength{\parskip}{0pt}}

\title{Deep Factorization Machines for Knowledge Tracing}

\author{Jill-J\^enn Vie \\
  RIKEN Center for Advanced Intelligence Project \\
  Nihonbashi 1-4-1, Mitsui Building 15F \\
  Chuo-ku, 103-0027 Tokyo, Japan \\
  {\tt vie@jill-jenn.net}}

\date{}

\begin{document}
\maketitle

\begin{abstract}
This paper introduces our solution to the 2018 Duolingo Shared Task on Second Language Acquisition Modeling (SLAM). We used deep factorization machines, a wide and deep learning model of pairwise relationships between users, items, skills, and other entities considered. Our solution (AUC 0.815) hopefully managed to beat the logistic regression baseline (AUC 0.774) but not the top performing model (AUC 0.861) and reveals interesting strategies to build upon item response theory models.
\end{abstract}

\hypertarget{introduction}{%
\section{Introduction}\label{introduction}}

Given the massive amount of data collected by online platforms, it is
natural to wonder how to use it to personalize learning. Students should
receive, based on their estimated knowledge, tailored exercises and
lessons, so they can be guided through databases of potentially millions
of exercises.

With this objective in mind, numerous models have been designed for
student modeling \citep{Desmarais2012}. Based on the outcomes of
students, one can infer the parameters of these so-called student
models, measure knowledge, and tailor instruction accordingly.

In the 2018 Duolingo Shared Task on Second Language Acquisition Modeling
\citep{slam18}, we had access to attempts of thousands of students over
sentences (composed of thousands of possible words, each of these being
labeled as correct or incorrect), and we had to predict whether a
student would write correctly or not the words of a new sentence.
Sentences were annotated with precious side information such as lexical,
morphological, or syntactic features. This problem is coined as
knowledge tracing \citep{corbett1994knowledge} or predicting student
performance \citep{minaei2003predicting} in the literature. In this
particular challenge, it is done at the word level.

In this paper, we explain the motivations that led us to our solution,
and show how our models handle typical models in educational data mining
as special cases. In Section 2, we show related work. In Section 3, we
present the existing model of DeepFM and clarify how it can be applied
for knowledge tracing, notably the SLAM task. In Section 4, we detail
the data preparation, in order to apply DeepFM. Finally, we expose our
results in Section 5 and further work in Section 6.

\hypertarget{related-work}{%
\section{Related Work}\label{related-work}}

Item Response Theory (IRT) models \citep{Hambleton1991} have been
extensively studied and deployed in many real-world applications such as
standardized tests (GMAT). They model the ability (level information) of
students, and diverse parameters of items (such as difficulty), and
involve many criteria for the selection of items to measure the ability
of examinees.

Related work in knowledge tracing consists in predicting the sequence of
outcomes for a given learner. Historically, Bayesian Knowledge Tracing
(BKT) modeled the learner as a Hidden Markov model
\citep{corbett1994knowledge}, but with the advent of deep learning, a
Deep Knowledge Tracing (DKT) model has been proposed
\citep{piech2015deep}, relying on long short-term memory
\citep{hochreiter1997long}. However, \citet{wilson2016back} have shown
that a simple variant of IRT could outperform DKT models.

All of these IRT, BKT or DKT models do not consider side information,
such as knowledge components, which is why new models naturally rose.
\citet{Vie2018} have used Bayesian factorization machines for knowledge
tracing, and recovered most student models as special cases.

Wide and deep learning models have been proposed by Google
\citep{cheng2016wide} to learn lower-order and higher-order features.
\citet{guo2017deepfm} have proposed a variant where they replace the
wide linear model by a factorization machine, and this is the best model
we got for the Shared Task challenge.

\hypertarget{deepfm-for-knowledge-tracing}{%
\section{DeepFM for knowledge
tracing}\label{deepfm-for-knowledge-tracing}}

We now introduce some vocabulary. We assume that our observed instances
can be described by \(C\) \emph{categories} of discrete or continuous
features (such as \texttt{user\_id}, \texttt{item\_id} or
\texttt{country}, but also \texttt{time}). \emph{Entities} denote
couples of categories and discrete values (such as \texttt{user=2},
\texttt{country=FR} or again \texttt{time} if the category is
continuous). We denote by \(N\) the number of possible entities, number
them from 1 to \(N\). The DeepFM model we are describing will learn an
embedding for each of those entities\footnote{The original DeepFM paper
  \citep{guo2017deepfm} chooses \emph{fields} and \emph{features} in
  lieu of \emph{categories} or \emph{entities}, but we prefer to use our
  own formulation \citep{Vie2018} because we usually agree with
  ourselves.}.

Each instance can be encoded as a sparse vector \(\bm{x}\) of size
\(N\): each component will be set at a certain value (for example, 1 if
the category of the corresponding entity is discrete, the value itself
if it is continuous, and 0 if the entity is not present in the
observation). For each instance, our model will output a probability
\(p(\bm{x}) = \psi(y_{FM} + y_{DNN})\), where \(\psi\) is a link
function such as the sigmoid \(\sigma\) or the cumulative distribution
function (CDF) \(\Phi\) of the standard normal distribution.

The DeepFM model is made of two components, the FM component and the
Deep component.

\hypertarget{fm-component}{%
\subsection{FM component}\label{fm-component}}

Given an embedding size \(d \in \mathbf{N}\), the output of a
factorization machine is the following:

\[ y_{FM} = \sum_{k = 1}^N w_k x_k + \sum_{1 \leq k < l \leq N} x_k x_l \langle \bm{v_k}, \bm{v_l} \rangle \]

The first term shows that a bias \(w_k \in \R\) is learned for each
entity \(k\). The second term models the pairwise interactions between
entities by learning a vector \(\bm{v_k} \in \R^d\) for each entity
\(k\).

\hypertarget{relation-to-existing-student-models}{%
\subsubsection{Relation to existing student
models}\label{relation-to-existing-student-models}}

If \(d = 0\) and \(\psi\) is the sigmoid function \(\sigma\),
\(p(\bm{x}) =\sigma(\langle \bm{w}, \bm{x} \rangle)\) and the FM
component behaves like logistic regression.

In particular, if there are two fields users (of \(n\) possible values)
and items, then each instance encoding \(\bm{x}_{ij}\) of user \(i\) and
item \(j\) is a concatenation of two one-hot vectors, and
\(p(\bm{x}_{ij}) = \sigma(w_i + w_{n + j}) = \sigma(\theta_i - d_j)\)
for appropriate values of \(w\), which means the Rasch model is
recovered.

As pointed out by \citet{slam18}, their baseline model is a logistic
regression with side information, which makes it similar to an additive
factor model. To see more connections between our FM component and
existing educational data mining models, see \citet{Vie2018}.

\hypertarget{deep-component}{%
\subsection{Deep component}\label{deep-component}}

The deep component is a \(L\)-layer feedforward neural network that
outputs:

\[ y_{DNN} = \ReLU(W^{(L)} a^{(L)} + b^{(L)}) \]

\noindent where each layer \(0 \leq \ell < L\) verifies:

\[ a^{(\ell + 1)} = \ReLU(W^{(\ell)} a^{(\ell)} + b^{(\ell)}) \]

\noindent for learned parameters \(W\), \(a\), \(b\) for each layer, and
the first layer is given by the corresponding \(\bm{v_{i_c}}\)
embeddings of the activated entities (the ones for each category
\(c = 1, \ldots, C\), which correspond to the nonzero entries of
\(\bm{x}\)):

\[ a^{0} = (\bm{v_{i_1}}, \ldots, \bm{v_{i_C}}). \]

In order to select the hyperparameters, we followed the instructions of
\citep{guo2017deepfm} and the default values of the available
implementation on GitHub\footnote{https://github.com/ChenglongChen/tensorflow-DeepFM}.

\hypertarget{training}{%
\subsection{Training}\label{training}}

Training is performed by minimizing the log loss of the output
probabilities compared to the true outcomes of the students over the
tokens. For all models trained, the optimizer was Adam
\citep{kingma2014adam}, with learning rate \(\gamma = 10^{-3}\) and
minibatches of size 1024.

\hypertarget{encoding-the-duolingo-dataset}{%
\section{Encoding the Duolingo
Dataset}\label{encoding-the-duolingo-dataset}}

\hypertarget{fundamental-discrete-categories}{%
\subsection{Fundamental, discrete
categories}\label{fundamental-discrete-categories}}

Fundamental categories (\texttt{\textless{}fundamental\textgreater{}})
refer to the features that have discrete values, such as \texttt{user}
(which refer to the user ID) or \texttt{countries} (which can be in a
many-to-many relationship).

\begin{multicols}{2}
\begin{itemize}
\tightlist
\item \verb+user+
\item \verb+token+
\item \verb+part_of_speech+
\item \verb+dependency_label+
\item \verb+exercise_index+
\item \verb+countries+
\item \verb+client+
\item \verb+session+
\item \verb+format+
\end{itemize}
\end{multicols}

\hypertarget{noisy-discrete-categories}{%
\subsection{Noisy discrete categories}\label{noisy-discrete-categories}}

Duolingo was providing the SyntaxNet features (morphosyntactic rules)
such as:

\begin{multicols}{2}
\begin{itemize}
\tightlist
\item \verb+Definite+
\item \verb+Gender+
\item \verb+Number+
\item \verb+fPOS+
\item \verb+Person+
\item \verb+PronType+
\item \verb+Mood+
\item \verb+Tense+
\item \verb+VerbForm+
\end{itemize}
\end{multicols}

We call them noisy (\texttt{\textless{}noisy\textgreater{}} below),
because they are the output of another algorithm. Also, not all of them
were specified, there were some missing entries.

\hypertarget{continuous-categories}{%
\subsection{Continuous categories}\label{continuous-categories}}

\begin{itemize}
\tightlist
\item
  \texttt{time} for answering the question
\item
  \texttt{days} since when the user subscribed the Duolingo platform.
\end{itemize}

\hypertarget{encoding}{%
\subsection{Encoding}\label{encoding}}

In the baseline model provided by Duolingo, all fundamental features
were encoded as a concatenation of \(n\)-hot encoders\footnote{For this
  reason, the continuous features could not be used for the baseline.}.
Then they used logistic regression and achieved AUC 0.772.

Here are the models we considered.

\begin{itemize}
\tightlist
\item
  IRT: \texttt{user} + \texttt{token}, \(d = 0\)
\item
  Logistic regression baseline:
  \texttt{\textless{}fundamental\textgreater{}}
\item
  Vanilla FM: \texttt{\textless{}fundamental\textgreater{}}
\item
  DeepFM: \texttt{\textless{}fundamental\textgreater{}}
\item
  DeepFM*:
  \texttt{\textless{}fundamental\textgreater{}\ +\ \textless{}noisy\textgreater{}\ +\ \textless{}continuous\textgreater{}}
\end{itemize}

The implementation of Deep Factorization Machines we used needed a
concatenation of one-hot encoders. So we picked the first country among
the list of countries for each instance. Also, it could not handle
missing entries, so for the noisy partial categories, we used a
\texttt{None} entity.

\hypertarget{results}{%
\section{Results}\label{results}}

We first tried different models on a validation set. All models were
trained using 500 epochs for the vanilla FM, or 100 epochs for DeepFM
with early stopping, and refit on the validation set.

\hypertarget{on-validation-set}{%
\subsection{On validation set}\label{on-validation-set}}

A vanilla FM was used considering \(\psi = \Phi\) the CDF of the
standard normal distribution as link function, like in the
implementation of\footnote{http://www.libfm.org}
\citep{rendle2012factorization}. Then, for our experiments, we used the
TensorFlow implementation of DeepFM provided by Alibaba on
GitHub\footnote{https://github.com/ChenglongChen/tensorflow-DeepFM}. Our
encoding is available on GitHub\footnote{https://github.com/jilljenn/ktm}.

\begin{table}
\begin{tabular}{ccccc} \toprule
& ACC & AUC & NLL & F1\\ \midrule
IRT + attempts & 0.833 & 0.739 & 0.411\\
Basic IRT & 0.838 & 0.752 & 0.399\\
LR baseline & 0.838 & 0.772 & 0.391 & 0.284\\
Vanilla FM & 0.824 & 0.773 & 0.414\\
DeepFM* & & 0.811 & & \textbf{0.382}\\
\textbf{DeepFM} & & \textbf{0.815} & & 0.329\\ \bottomrule
\end{tabular}
\caption{Performance of all tested algorithms on the \texttt{en\_es} dataset.}
\end{table}

Vanilla FM had comparable performance of the LR baseline. It agrees with
the findings of \citet{Vie2018} that a bigger dimension may not
necessarily help.

\hypertarget{on-test-set}{%
\subsection{On test set}\label{on-test-set}}

The DeepFM model managed to improve the baseline by 3 points AUC. We got
AUC 0.815, while the top performing solution had AUC 0.861.

Our best performing model was DeepFM: using only the discrete features,
train a model of latent embedding size 10 during a fixed number of
epochs (50). DeepFM* using all features was slightly worse.

\hypertarget{further-work}{%
\section{Further Work}\label{further-work}}

We could embed the dependency graph provided by Duolingo in the encoding
of the vanilla FM.

Ensemble methods such as \texttt{xgboost} \citep{chen2016xgboost} could
be considered, as typically encountered in challenges.

Here we want to combine information of the student which is quite poor
(almost only their outcomes), compared to the knowledge of tokens
(syntactic trees, or word2vec, etc.). This is why we could use extra
embeddings, such as a LSTM encoding of the sentence as feature for the
token.

The performance of DeepFM* that was using all features was slightly
worse than DeepFM that was limited to the fundamental features. We might
mitigate this problem by using a field-aware factorization machine
\citep{juan2016field} that learns a parameter per category of feature in
order to draw more importance on some category (such as \texttt{user})
than others (such as \texttt{date}).

\hypertarget{conclusion}{%
\section{Conclusion}\label{conclusion}}

In this paper, we showed how to use deep factorization machines for
knowledge tracing. Our findings show interesting combinations of
features, together with embeddings provided by deep neural networks. In
some way, it shows how to learn dense embeddings from the sparse
features typically encountered in learning platforms.

\bibliography{biblio}
\bibliographystyle{acl_natbib}

\end{document}